\documentclass{elsart}
\usepackage{amssymb}
\newcommand{\bfm}[1]{\mbox{\boldmath${#1}$}}
\begin{document}
\begin{frontmatter}
\title{Lesche Stability of $\bf\kappa$-Entropy}
\author[INFM]{G. Kaniadakis}
\author[INFN,INFM]{A.M. Scarfone}
\address[INFM]{Istituto Nazionale di
Fisica della Materia and Dipartimento di Fisica,\\ Politecnico di
Torino, Corso Duca degli Abruzzi 24, 10129 Torino, Italy}
\address[INFN]{Istituto Nazionale di
Fisica Nucleare and Dipartimento di Fisica,\\ Universit\`a di
Cagliari, Cittadella Universitaria, 09042 Monserrato, Cagliari,
Italy}
\date{\today}

\begin {abstract}
The Lesche stability condition for the Shannon entropy [B.
Lesche, J. Stat. Phys. \bfm{27}, 419 (1982)], represents a
fundamental test, for its experimental robustness, for systems
obeying the Maxwell-Boltzmann statistical mechanics. Of course,
this stability condition must be satisfied by any entropic
functional candidate to generate non-conventional statistical
mechanics. In the present effort we show that the
$\kappa$-entropy, recently introduced in literature [G.
Kaniadakis, Phys. Rev. E \bfm{66}, 056125 (2002)], satisfies the
Lesche stability condition.
\end {abstract}

\end{frontmatter}
\maketitle

It is widely accepted that the Boltzmann-Gibbs statistical
mechanics, while succeeding to describe systems in a stationary
state characterized by an ergodicity consistent with the thermal
equilibrium, fails to depict of the statistical properties of
anomalous systems as found in surface growth, anomalous diffusion
in porous media, gravitation, L\'evy flights, fractals,
turbulence physics, and economics
\cite{Abe0,Kaniadakis0}.\\
Generalized statistical mechanics have been developed with the
purpose of dealing with anomalous systems by means of the same
mathematical tools used in the conventional statistical mechanics.
This can be accomplished, in the case of trace-form entropies,
with the introduction of an appropriate deformed logarithm.
Examples of trace-form entropies are the Tsallis entropy
\cite{Tsallis} and the Abe entropy \cite{Abe}. In
\cite{Naudts1,Naudts2} some general properties of the deformed
logarithms, leading to a generalization of the Boltzmann-Gibbs
distributions, have been studied.\\ Clearly, there are also
entropies which are not of trace-form, e.g. the R\'enyi entropy
\cite{Renyi}, the Landsberg-Vedral entropy \cite{Landsberg} and
the escort entropy \cite{Tsallis4}. In these cases the definition
of these generalized entropies does not necessitate the
preliminary introduction of a deformed logarithm.

Recently, a non-conventional statistical mechanics has been
introduced in \cite{Kaniadakis1,Kaniadakis2} which preserves the
epistemological and thermodynamics structure of the standard
Boltzmann-Shannon-Gibbs theory. The proposed theory is based on
the following trace-form entropy ($k_{_{\rm B}}=1$)
\begin{equation}
S_{_\kappa}(p)=-\sum_{i=1}^N\,p_{_i}\,\ln_{_{\{{\scriptstyle
\kappa}\}}}(p_{_i}) \ ,\label{kentropy}
\end{equation}
being $p\equiv\{p_{_i},\,i=1,\,\cdots,\,N\}$ a discrete
probability distribution normalized as
\begin{equation}
\sum_{i=1}^Np_{_i}=1 \ ,\label{norm}
\end{equation}
and the $\kappa$-logarithm is defined by
\begin{equation}
\ln_{_{\{{\scriptstyle
\kappa}\}}}(x)=\frac{x^\kappa-x^{-\kappa}}{2\,\kappa} \
.\label{klog}
\end{equation}
The $\kappa$-logarithm is a monotonically increasing and concave
function for $\kappa\in(-1,\,1)$, being $d\ln_{_{\{{\scriptstyle
\kappa}\}}}(x)/dx>0$ and $d^2\ln_{_{\{{\scriptstyle
\kappa}\}}}(x)/dx^2<0$, and satisfies the relation
\begin{equation}
\ln_{_{\{{\scriptstyle \kappa}\}}}(x)=-\ln_{_{\{{\scriptstyle
\kappa}\}}}\left(\frac{1}{x}\right) \ .
\end{equation}
In the limit of $\kappa\rightarrow0$ Eq. (\ref{klog}) reduces to
the standard logarithm and consequently Eq. (\ref{kentropy})
converges to the well-known Boltzmann-Shannon-Gibbs entropy.\\
The inverse function of the $\kappa$-logarithm, namely the
$\kappa$-exponential, is defined as
\begin{equation}
\exp_{_{\{{\scriptstyle
\kappa}\}}}(x)=\left(\sqrt{1+\kappa^2\,x^2}+\kappa\,x\right)^{1/\kappa}
\ .\label{kexp}
\end{equation}
It is a positive, monotonically increasing and convex function
which reduces to the standard exponential for $\kappa\rightarrow0$
and satisfies the relation
\begin{equation}
\exp_{_{\{{\scriptstyle \kappa}\}}}(x)\,\exp_{_{\{{\scriptstyle
\kappa}\}}}(-x)=1 \ .\label{invexp}
\end{equation}
After maximizing the $\kappa$-entropy given by Eq.
(\ref{kentropy}) with the constraints (\ref{norm}) and
\begin{equation}
\sum_{i=1}^N E_{_i}\,p_{_i}=\langle E\rangle \ ,\label{emean}
\end{equation}
being $E_{_i}$ the energy of the $i$-th level with probability of
occupation  $p_{_i}$, we obtain the probability distribution
\begin{equation}
p_{_i}=\alpha\,\exp_{_{\{{\scriptstyle
\kappa}\}}}\left(-\frac{\beta}{\lambda}\,(E_{_i}-\mu)\right) \ ,
\end{equation}
where $\beta=1/T$ and $-\beta\,\mu$ are the Lagrange multipliers
associated to the constraints given by Eq.s (\ref{norm}) and
(\ref{emean}), while the constants $\alpha$ and $\lambda$ depend
on the parameter $\kappa$ and are given respectively by $
\alpha=[(1-\kappa)/(1+\kappa)]^{1/2\,\kappa}$ and
$\lambda=\sqrt{1-\kappa^2}$.

We recall now briefly the physical motivations which justify  the
introduction of $\kappa$-statistical mechanics. For a more
detailed discussion we refer the reader to \cite{Kaniadakis1}.\\
First we remark that the $\kappa$-exponential and the
$\kappa$-logarithm have the two following mathematical properties
\begin{equation}
\exp_{_{\{{\scriptstyle \kappa}\}}}(x)\,\exp_{_{\{{\scriptstyle
\kappa}\}}}(y)=\exp_{_{\{{\scriptstyle
\kappa}\}}}(x\oplus\!\!\!\!\!^{^{\scriptstyle \kappa}}\,\, y) \ ,
\end{equation}
\begin{equation}
\ln_{_{\{{\scriptstyle \kappa}\}}} (xy)=\ln_{_{\{{\scriptstyle
\kappa}\}}} (x)\oplus\!\!\!\!\!^{^{\scriptstyle \kappa}}\,\,
\ln_{_{\{{\scriptstyle \kappa}\}}} (y) \ ,
\end{equation}
where the $\kappa$-sum $x\oplus\!\!\!\!\!^{^{\scriptstyle
\kappa}}\,\, y$ is defined as
\begin{equation}
x\oplus\!\!\!\!\!^{^{\scriptstyle \kappa}}\,\,
y=x\,\sqrt{1+\kappa^2\,y^2}+y\,\sqrt{1+\kappa^2\,x^2_{ }} \
.\label{ksum}
\end{equation}
The above properties in the limit $\kappa \rightarrow 0$ reduce
to the well-known relations $\exp (x)\,\exp (y) = \exp (x+y)$ and
$\ln (xy)=\ln (x) + \ln (y)$ being
$x\oplus\!\!\!\!^{^{\scriptstyle 0}}\,\, y = x+y$. Of course the
$\kappa$-sum plays a central role in the construction of the
theory. In Ref. \cite{Kaniadakis1,Kaniadakis2} it has been shown
that the algebra constructed starting from the $\kappa$-sum
induces a $\kappa$-analysis. It results that the
$\kappa$-derivative admits as eigenstate  the
$\kappa$-exponential just as the ordinary exponential emerges as
eigenstate of the ordinary derivative. At this point is evident
that the physical motivation of the theory is strictly related
with the physical mechanism originating the $\kappa$-deformation.
In Ref. \cite{Kaniadakis1} it has been shown that the
$\kappa$-sum emerges naturally within the Einstein special
relativity. More precisely the $\kappa$-sum is related to the
relativistic sum of the velocities
\begin{equation}
v_{_1}\oplus\!\!^{^{\scriptstyle
c}}\,\,v_{_2}=\frac{v_{_1}+v_{_2}}{1+v_{_1}\,v_{_2}/c^2} \ ,
\end{equation}
through the relation
\begin{equation}
p(v_{_1})\oplus\!\!\!\!\!^{^{\scriptstyle
\kappa}}\,\,p(v_{_2})=p(v_{_1}\oplus\!\!^{^{\scriptstyle
c}}\,\,v_{_2}) \ ,
\end{equation}
with $\kappa=1/m\,c$, where
\begin{equation}
p(v_{_i})=\frac{m\,v_{_i}}{\sqrt{1-v_{_i}^2/c^2}} \ ,
\end{equation}
is the relativistic momentum of a particle of rest mass $m$. Only
in the classic limit $c \rightarrow \infty$ the parameter
$\kappa$ approaches zero and both the $\kappa$-sum and the
relativist sum of the velocities reduce to the ordinary sum. Thus
the $\kappa$-deformation is originated from the finite value of
light speed and results to be a purely relativistic effect.\\
Clearly when we consider a statistical system of relativistic
particles the $\kappa$-sum continues to play an important role.
Due to the finite values of the light speed $c\not=\infty$, any
signal and then any information propagates with a finite velocity
in the system and therefore it results $\kappa\not=0$.\\
The cosmic rays represents the most important example of
relativistic statistical system which manifestly violates the
Boltzmann statistics. In Ref. \cite{Kaniadakis1} it has been
shown that the $\kappa$-statistics can predict very well the
experimental cosmic ray spectrum. This is an important test for
the theory because the cosmic rays spectrum has a very large
extension (13 decades in energy and 33 decades in flux).\\
Let us consider now statistical systems (physical, natural,
economical, etc.) in which is involved a limiting quantity, like
the light speed in the relativistic particle system. For these
systems where the information propagates with finite velocity, it
is reasonable suppose that the $\kappa$-deformation can appear,
so that the $\kappa$-statistics results to be the most
appropriate theory to describe these systems.

An important question is if the $\kappa$-entropy can be
associated to a physical system in an experimentally detectable
state. The problem can be stated in the following way. Let
${\mathcal O}(\{x\})$ be a physical observable, depending on a
set of parameters $\{x\}\equiv\{x_{_i},\,i=1,\,\cdots,\,N\}$ with
${\rm sup}[{\mathcal O}(\{x\})]$ its maximal value achieved while
varying $\{x\}$. ${\mathcal O}(\{x\})$ is stable if, under a small
change of the parameters $\{x\}\rightarrow\{x^\prime\}$, with
$|\{x^\prime\}-\{x\}|<\delta$, the relative discrepancy
\begin{equation}
\Delta=\frac{{\mathcal O}(\{x^\prime\})-{\mathcal O}(\{x\})}{{\rm
sup}[{\mathcal O}(\{x\})]} \ ,
\end{equation}
can be reduced to zero.\\
This question has been considered, for the first time, by Lesche
\cite{Lesche} in 1982 both for the Boltzmann-Shannon-Gibbs
entropy and for the R\'enyi entropy, showing that the former is
stable while the latter is unstable. Successively, Abe
\cite{Abe1} showed that the Tsallis entropy is also consistent
with the stability condition, while the Landsberg-Vedral entropy
does not satisfy the stability condition. In \cite{Tsallis3} it
has been shown that also the escort entropy does not satisfy the
experimental robustness criterium. Finally, in \cite{Naudts2} it
has been considered the problem of Lesche stability condition from
an arbitrary distribution function.\\
We remark that the Lesche inequality represents a condition for
the stability of a system under particular sorts of perturbations.
Clearly, the R\'enyi entropy, the Landsberg-Vedral entropy as well
as the escort entropy, which fail to satisfy the Lesche stability
condition, are worthy of investigation because present many other
interesting properties.\\
The purpose of the present contribution is to show that the
$\kappa$-entropy defined by Eq. (\ref{kentropy}) satisfies the
Lesche stability
condition and thus it can represent a well defined physical observable.\\

Let us begin by introducing the following auxiliary function
\begin{equation}
A_{_\kappa}(p,\,s)=\sum_{i=1}^N\left[p_{_i}-\alpha\,\exp_{_{\{{\scriptstyle
\kappa}\}}}(-s)\right]_+ \ ,\label{A}
\end{equation}
being $s$ a real positive parameter. The function $[x]_+$ is
defined in terms of the Heaviside unit step function $\theta(x)$
through $[x]_+=x\,\theta(x)$ so that it results $[x]_+=x$ for
$x>0$ and
$[x]_+=0$ for $x\leq0$.\\
We remark that the function $A_{_\kappa}(p,\,s)$ has been
considered already in the literature. Originally, in Ref.
\cite{Lesche} $A_{_\kappa}(p,\,s)$ is defined using the
$\exp(-s)$ at the place of $\alpha\,\exp_{_{\{{\scriptstyle
\kappa}\}}}(-s)$. In Ref. \cite{Abe1} the function
$A_{_\kappa}(p,\,s)$ is defined starting from the Tsallis
exponential $\exp_{_q}(s)$. Very recently, in Ref. \cite{Abe2},
it was considered a very large family of functions having the
structure of (\ref{A}) where at the place of
$\alpha\,\exp_{_{\{{\scriptstyle \kappa}\}}}(-s)$ appears an
arbitrary positive real function $f(s)$.\\
First we observe that from the relation $|[x]_+-[y]_+|\leq|x-y|$
it follows that $\forall s$ holds the inequality
\begin{equation}
\Big|
A_{_\kappa}(p,\,s)-A_{_\kappa}(q,\,s)\Big|\leq|\!|p-q|\!|_{_1} \
,\label{rr2}
\end{equation}
being
\begin{equation}
|\!|p-q|\!|_{_1}=\sum_{i=1}^N|p_{_i}-q_{_i}| \ .
\end{equation}
Second, following the procedure described in \cite{Lesche,Abe1}
we have that from the definition of $A_{_\kappa}(p,\,s)$ follows
\begin{equation}
\sum_{i=1}^N \left[p_{_i}-\alpha\,\exp_{_{\{{\scriptstyle
\kappa}\}}}(-s)\right]_+<\sum_{i=1}^N p_{_i} \ ,\label{p0}
\end{equation}
which implies $A_{_\kappa}(p,\,s)<1$,
being $\{p_{_i}\}$ a probability distribution normalized to one.\\
On the other hand, by making use of the relation
$\sum_i[x_{_i}]_+\geq[\sum_i x_{_i}]_+$ we have
\begin{eqnarray}
\nonumber\hspace{10mm}\sum_{i=1}^N
\left[p_{_i}-\alpha\,\exp_{_{\{{\scriptstyle
\kappa}\}}}(-s)\right]_+&\geq&\left[\sum_{i=1}^N
\left(p_{_i}-\alpha\,\exp_{_{\{{\scriptstyle
\kappa}\}}}(-s)\right)\right]_+\\
&=&\left[1-\alpha\,N\,\exp_{_{\{{\scriptstyle
\kappa}\}}}(-s)\right]_+ \ ,\label{p1}
\end{eqnarray}
and posing $s\geq\ln_{_{\{{\scriptstyle \kappa}\}}}(N)$, being
$\alpha<1$, we can drop the notation $[\cdot\hspace{.1mm}]_+$ in
the r.h.s of Eq. (\ref{p1}) obtaining
\begin{equation}
A_{_\kappa}(p,\,s)\geq1-\alpha\,N\,\exp_{_{\{{\scriptstyle
\kappa}\}}}(-s) \ .\label{d2}
\end{equation}
Combining Eq.s (\ref{p0}) and (\ref{d2}) we obtain
\begin{equation}
1-\alpha\,N\,\exp_{_{\{{\scriptstyle \kappa}\}}}(-s)\leq
A_{_\kappa}(p,\,s)<1 \ .\label{rang1}
\end{equation}
Of course, this condition remains true for any other discrete
probability distribution $\{q_{_i}\}$. Then, after changing the
sign of Eq. (\ref{rang1}), we obtain
\begin{equation}
-1<-A_{_\kappa}(q,\,s)\leq-1+\alpha\,N\,\exp_{_{\{{\scriptstyle
\kappa}\}}}(-s) \ .\label{rang2}
\end{equation}
Summing Eq.s (\ref{rang1}) and (\ref{rang2}) follows
\begin{equation}
-\alpha\,N\,\exp_{_{\{{\scriptstyle
\kappa}\}}}(-s)<A_{_\kappa}(p,\,s)-A_{_\kappa}(q,\,s)<\alpha\,N\,\exp_{_{\{{\scriptstyle
\kappa}\}}}(-s) \ ,
\end{equation}
and than
\begin{equation}
\Big|
A_{_\kappa}(p,\,s)-A_{_\kappa}(q,\,s)\Big|<\alpha\,N\,\exp_{_{\{{\scriptstyle
\kappa}\}}}(-s) ,\label{rr1}
\end{equation}
holding for $s\geq\ln_{_{\{{\scriptstyle \kappa}\}}}(N)$. Eq.s
(\ref{rr2}) and (\ref{rr1}) express two very important properties
of the function
$A_{_\kappa}(p,\,s)$ which will be used in the following.\\
Now we show that the entropy $S_{_\kappa}(p)$ can be expressed in
terms of the function $A_{_\kappa}(p,\,s)$. In fact, from the
definition (\ref{kentropy}) and taking into account the identity
$\alpha^\kappa\,(1+\kappa)=\lambda$, we have
\begin{eqnarray}
\nonumber\hspace{10mm} S_{_\kappa}(p)&=&-\sum_{i=1}^N
p_{_i}\left(\frac{p_{_i}^\kappa-p_{_i}^{-\kappa}}{2\,\kappa}\right)\\
\nonumber &=&-\frac{\lambda}{2\,\kappa}\sum_{i=1}^N p_{_i}\,
\left[{1\over1+\kappa}\,\left(\frac{p_{_i}}{\alpha}\right)^\kappa-
{1\over1-\kappa}\,\left(\frac{p_{_i}}{\alpha}\right)^{-\kappa}\right]\\
\nonumber &=&-\frac{\lambda}{2\,\kappa}\sum_{i=1}^N
p_{_i}\,\left[\left(\frac{p_{_i}}{\alpha}\right)^\kappa-\left(\frac{p_{_i}}{\alpha}\right)^{-\kappa}\right]\\
\nonumber&+&
{\lambda\over2}\sum_{i=1}^Np_{_i}\,\left[{1\over1+\kappa}\left(\frac{p_{_i}}{\alpha}\right)^\kappa+
{1\over1-\kappa}\left(\frac{p_{_i}}{\alpha}\right)^{-\kappa}\right]\\
&=&\lambda\sum_{i=1}^N\int\limits_0\limits^{L_i} p_{_i}\,
ds+\alpha\,\lambda\,\sum_{i=1}^N\int\limits_0\limits^{p_{_i}/\alpha}
x\,\frac{d}{dx}\ln_{_{\{{\scriptstyle \kappa}\}}}(x)\,dx\ \ ,\
\end{eqnarray}
where $L_{_i}=-\ln_{_{\{{\scriptstyle \kappa}\}}}(p_{_i}/\alpha)$.
After introducing the substitution $s=-\ln_{_{\{{\scriptstyle
\kappa}\}}}(x)$ we obtain
\begin{eqnarray}
\nonumber
\hspace{15mm}S_{_\kappa}(p)&=&\lambda\sum_{i=1}^N\int\limits_0\limits^{L_i}
p_{_i}\, ds+\lambda\sum_{i=1}^N\int\limits_{L_i}\limits^\infty
\alpha\,\exp_{_{\{{\scriptstyle \kappa}\}}}(-s)\,ds\\
&=&\lambda\,\int\limits_0\limits^\infty
ds-\lambda\sum_{i=1}^N\int\limits_{L_i}\limits^\infty
\left[p_{_i}-\alpha\,\exp_{_{\{{\scriptstyle
\kappa}\}}}(-s)\right]\,ds \ ,
\end{eqnarray}
and finally we can write the entropy $S_{_\kappa}(p)$ in terms of
the function $A_{_\kappa}(p,\,s)$ as follows
\begin{equation}
S_{_\kappa}(p)=\lambda\int\limits_{-1/\lambda}\limits^\infty
\left[1-A_{_\kappa}(p,\,s)\right]\,ds-1 \ .\label{kent}
\end{equation}
Using Eq. (\ref{kent}) the difference of $\kappa$-entropy for two
probability distributions, namely $p=\{p_{_i}\}$ and
$q=\{q_{_i}\}$ can be written
\begin{eqnarray}
\nonumber\hspace{15mm}\Big| S_{_\kappa}(p)-S_{_\kappa}(q)\Big|
&=&\lambda\,\Bigg|\int\limits_{-1/\lambda}\limits^\infty\left[
A_{_\kappa}(p,\,s)-A_{_\kappa}(q,\,s)\right]\,ds\Bigg|
\\
&\leq& \lambda\int\limits_{-1/\lambda}\limits^\infty
\Big|A_{_\kappa}(p,\,s)-A_{_\kappa}(q,\,s)\Big|\,ds \ . \label{e1}
\end{eqnarray}
After splitting the integration interval in two parts, namely
$[{-1/\lambda},\,+\infty)=[{-1/\lambda},\,\ell]\bigcup[\ell,\,+\infty)$,
Eq. (\ref{e1}) becomes
\begin{eqnarray}
\nonumber\hspace{15mm} \Big|
S_{_\kappa}(p)-S_{_\kappa}(q)\Big|&&\leq\lambda\int\limits_{-1/\lambda}\limits^\ell\Big|
A_{_\kappa}(p,\,s)-A_{_\kappa}(q,\,s)\Big|\,ds
\\
&& +\lambda\int\limits_\ell\limits^\infty\Big|
A_{_\kappa}(p,\,s)-A_{_\kappa}(q,\,s)\Big|\,ds \ , \label{s1}
\end{eqnarray}
where the splitting point $\ell$ will be defined below. Now, using
inequality (\ref{rr2}) in the first integral of Eq. (\ref{s1}) and
inequality (\ref{rr1}) in the second integral of Eq. (\ref{s1}),
we obtain:
\begin{equation}
\Big| S_{_\kappa}(p)-S_{_\kappa}(q)\Big|\leq\lambda\,{\mathcal
G}_{_\kappa}(\ell) \ ,\label{s3}
\end{equation}
where the function ${\mathcal G}_{_\kappa}(\ell)$ is defined by
\begin{equation}
{\mathcal
G}_{_\kappa}(\ell)=\int\limits_{-1/\lambda}\limits^\ell|\!|p-q|\!|_{_1}\,ds+
N\,\alpha\int\limits_\ell\limits^\infty \exp_{_{\{{\scriptstyle
\kappa}\}}}(-s)\,ds \ .\label{s0}
\end{equation}
After performing the integrations, Eq. (\ref{s0}) can be written
as
\begin{eqnarray}
\nonumber \!\!\!\!\!\!\!\!{\mathcal
G}_{_\kappa}(\ell)=\alpha\,\frac{N}{2}\left\{\frac{\left[
\exp_{_{\{{\scriptstyle
\kappa}\}}}(-\ell)\right]^{1+\kappa}}{1+\kappa}+\frac{\left[
\exp_{_{\{{\scriptstyle
\kappa}\}}}(-\ell)\right]^{1-\kappa}}{1-\kappa}\right\}
+|\!|p-q|\!|_{_1}\left(\ell+{1\over\lambda}\right) \ .\\
\label{s6}
\end{eqnarray}
Let us choose for the parameter $\ell$ the value $\ell_0$ which
minimizes the function ${\mathcal G}_{_\kappa}(\ell)$, therefore
after posing
\begin{equation}
\frac{d\,{\mathcal
G}_{_\kappa}(\ell)}{d\,\ell}\Big|_{\ell=\ell_0}=0 \ ,\label{dif}
\end{equation}
we obtain
\begin{equation}
\ell_0=\ln_{_{\{{\scriptstyle
\kappa}\}}}\left(\frac{\alpha\,N}{|\!|p-q|\!|_{_1}}\right) \
.\label{sol}
\end{equation}
The condition $s\geq\ln_{_{\{{\scriptstyle \kappa}\}}}(N)$,
holding for Eq. (\ref{rr1}), now requires
$\ell_0\geq\ln_{_{\{{\scriptstyle \kappa}\}}}(N)$ and then from
Eq. (\ref{sol}) we obtain
\begin{equation}
|\!|p-q|\!|_{_1}\leq\alpha \ .\label{con}
\end{equation}
By posing $\ell=\ell_0$ in the expression of ${\mathcal
G}_{_\kappa}(\ell)$, the inequality (\ref{s3}) becomes
\begin{eqnarray}
\nonumber\hspace{5mm}\Big|
S_{_\kappa}(p)-S_{_\kappa}(q)\Big|&\leq&|\!|p-q|\!|_{_1}\,
\left[{1\over2}\left( \frac{|\!|p-q|\!|_{_1}}{N}\right)^\kappa
+{1\over2}\left(
\frac{|\!|p-q|\!|_{_1}}{N}\right)^{-\kappa}\right.\\
&-&\left.\lambda \,\ln_{_{\{{\scriptstyle
\kappa}\}}}\left(\frac{|\!|p-q|\!|_{_1}}{\alpha\,N}\right)+1
\right] \ ,\label{s4}
\end{eqnarray}
and can be written in the compact form
\begin{equation}
\Big|
S_{_\kappa}(p)-S_{_\kappa}(q)\Big|\leq|\!|p-q|\!|_{_1}\,\left[
\ln_{_{\{{\scriptstyle
\kappa}\}}}\left(\frac{N}{|\!|p-q|\!|_{_1}}\right)+1\right] .
\label{s5}
\end{equation}
At this point we observe that $f(x)=-x\,\ln_{_{\{{\scriptstyle
\kappa}\}}}(x)$ is a positive and increasing function in the
interval $[0,\,\alpha]$. Furthermore we pose
$|\!|p-q|\!|_{_1}<\delta\leq\alpha$, consistently with Eq.
(\ref{con}).\\
By introducing the maximum value of $S_{_\kappa}(p)$ corresponding
to the uniform distribution
$p\equiv\{p_{_i}=1/N,\,\,i=1,\,\cdots,\,N\}$
\begin{equation}
S_{_\kappa}^{\rm max}=\ln_{_{\{{\scriptstyle \kappa}\}}}(N) \
,\label{Boltzmann}
\end{equation}
from Eq. (\ref{s5}) we can evaluate the entropy relative
discrepancy as
\begin{equation}
\Bigg|\frac{S_{_\kappa}(p)-S_{_\kappa}(q)}{S^{\rm
max}_{_\kappa}}\Bigg|\!\leq\!
\frac{\delta^{1-\kappa}}{1\!-\!N^{-2\,\kappa}}
-\frac{\delta^{1+\kappa}}{N^{2\,\kappa}\!-\!1}\!+\!\frac{2\,\kappa\,\delta}{N^\kappa\!-\!N^{-\kappa}}
\ .\label{erd1}
\end{equation}
Now, as customary, we evaluate the inequality given by Eq.
(\ref{erd1}) in the thermodynamic limit $N\rightarrow\infty$ and
obtain
\begin{equation}
\Bigg|\frac{S_{_\kappa}(p)-S_{_\kappa}(q)}{S^{\rm
max}_{_\kappa}}\Bigg|\leq\delta^{1-|\kappa|} \ .\label{final}
\end{equation}
Finally, by introducing in Eq. (\ref{final}) the positive quantity
$\epsilon=\delta^{1-|\kappa|}$, we obtain the Lesche inequality
\begin{equation}
\Bigg|\frac{S_{_\kappa}(p)-S_{_\kappa}(q)}{S^{\rm
max}_{_\kappa}}\Bigg|\leq\epsilon \ .\label{final1}
\end{equation}
Clearly in Eq. (\ref{final1}), $\epsilon$ is a continuous function
of $\delta$, approaching $0$ for $\delta\rightarrow0$ when
$-1<\kappa<1$ and therefore the $\kappa$-entropy $S_{_\kappa}$
satisfies the stability condition.


\end{document}